\documentstyle[aps,epsfig,multicol,fancyheadings,amssymb]{revtex}
\begin{document}

\title{Percolation in three-dimensional random field Ising magnets}

\author{E.\ T.\ Sepp\"al\"a,$^{1,2}$ A.\ M.\ Pulkkinen,$^2$ and
M.\ J.\ Alava$^{2,3}$}

\address{$^1$Lawrence Livermore National Laboratory, 7000 East Avenue, 
L-415, Livermore, CA 94550, U.S.A.}
\address{$^2$Helsinki University of Technology, Laboratory of Physics, 
P.O.Box 1100, FIN-02015 HUT, Finland}
\address{$^3$NORDITA, Blegdamsvej 17, DK-2100 Copenhagen, Denmark} 
\date{\today}

\maketitle

\begin{abstract}
The structure of the three-dimensional random field Ising magnet is
studied by ground state calculations. We investigate the percolation
of the minority spin orientation in the paramagnetic phase above the
bulk phase transition, located at $[\Delta/J]_c \simeq 2.27$, where
$\Delta$ is the standard deviation of the Gaussian random fields 
($J=1$).  With an external field $H$ there is a disorder
strength dependent critical field $\pm H_c(\Delta)$ for the down (or
up) spin spanning.  The percolation transition is in the standard
percolation universality class. $H_c \sim (\Delta -\Delta_p)^\delta$,
where $\Delta_p = 2.43 \pm 0.01$ and $\delta=1.31 \pm 0.03$, implying
a critical line for $\Delta_c < \Delta \leq \Delta_p$.  When,
with zero external field, $\Delta$ is decreased from a large value 
there is a transition from the simultaneous up and down spin spanning,
with probability $\Pi_{\uparrow \downarrow}=1.00$ to 
$\Pi_{\uparrow \downarrow}=0$. This is located at
$\Delta =2.32 \pm 0.01$, i.e., above $\Delta_c$.  The
spanning cluster has the fractal dimension of standard percolation
$D_f = 2.53$ at $H=H_c(\Delta)$. We provide evidence that this is
asymptotically true even at $H=0$ for $\Delta_c < \Delta \leq
\Delta_p$ beyond a crossover scale that diverges as $\Delta_c$ is
approached from above. Percolation implies extra finite size effects
in the ground states of the 3D RFIM.
\end{abstract}

\noindent {\it PACS \# \ \ 05.50.+q, 75.60.Ch, 75.50.Lk, 64.60.Ak}

\begin{multicols}{2}[]
\narrowtext

\section{Introduction}\label{intro}

The random-field Ising model (RFIM) is one of the most basic models
for random systems~\cite{Imr75,Young97,belangeryoung91}. Its beauty is
that the mixture of random fields and the standard Ising model creates
rich physics and leaves many still un-answered problems. By now it is
known that three dimensions (3D) is the corner stone of the model,
since it presents a phase transition where the randomness proves to be
a relevant perturbation to the pure 3D Ising model. For the last fifteen
years, since the method-wise seminal paper by
Ogielski~\cite{ogielski}, the studies of the transition have centered
around zero-temperature ground state computations because the
temperature is due to renormalization group arguments believed to be
an (perhaps dangerously so) irrelevant variable.

Many such works exist so far, the most recent and comprehensive being
due to Middleton and Fisher~\cite{middlefisher}. In spite of all the
effort many uncertainties remain concerning the nature of the phase
transition.  The question is if the transition is of the second-order,
of traditional first-order type, or finally some other kind of
discontinuous transition. The order-parameter exponent, $\beta$, may
have a finite value or it can be equal to zero
\cite{middlefisher,heikopeter,anglesourlas,hartmannowak,sourlas,billdave,hartmannyoung,machtaetal}. Its very small value makes it
unlikely that insight is obtained in the near future, in spite of the
fact that the optimization algorithms used can at best scale almost
linearly with the number of spins in the system.  Moreover a
controversy exists with regards to the role of disorder: the available
simulations are not able to settle the question whether the critical
exponents depend on the particular choice of the distribution for the
random fields, analogously to the mean-field theory of the RFIM where
binary ($\pm h$) disorder results in a first-order transition and
Gaussian (see below) in a second-order one~\cite{meinke}.

In this paper we focus on a novel aspect of the three-dimensional
RFIM, namely percolation~\cite{Amnon}.  The goal is to explore 
percolation critical phenomenona in the 3D RFIM.  The work is an
extension to our studies of percolation in two-dimensional
RFIM~\cite{RF2d_perk}. In the traditional 3D Ising model, without
disorder, the percolation behavior in an applied field and its
consequences, as whether the phase transition critical exponents would
be affected by the percolation criticality, are known since long as
the ``Trouble with Kert\'esz''~\cite{Amnon,kertesz}. This problem was
solved by Wang~\cite{wang} by studying Fortuin-Kasteleyn or
Coniglio-Klein~\cite{Fortuin,Coniglio} clusters using so called
ghost-spins. In the RFIM the situation is different in that at small
temperatures one has a {\em non-zero} spin-spin overlap $q$ with the
ground state: thus the existence of a ground state percolation
transition (even without an external field) implies measurable
consequences even at finite temperatures. It also complicates the
phase diagram by its existence.

There is one fundamental difference between two and three dimensions
(besides the fact that there is no phase transition in two dimensions,
and hence there systems are always paramagnetic). In two dimensional
square lattices the critical percolation site-occupation probability
is 0.592746, i.e., above one half, and in three dimensional cubic
lattices well below one half, 0.3116. Therefore in three dimensions,
deep in the paramagnetic phase, both of the spin orientations should span
the system (this has been noted by Esser {\it et al.} to be true for
the RFIM, see \cite{Esser}).  Thus introducing an external field in
paramagnetic systems leads in two dimensions to the percolation of the
spin direction parallel with the external field. In three dimensions,
on the other hand, the external field destroys the spanning property
of the spin orientation opposite to the external field.

Consequences of the percolation type of order at the paramagnetic
phase are many-fold. There are experimentally accessible random field
magnets, so called diluted antiferromagnets in an external field
(DAFF)~\cite{Belanger97} in which the percolation order could be
seen, should it exist for zero external fields. 
It is already known that the percolation of the diluted atoms
has a strong contribution to the behavior of the structure factor
line-shapes of the 3D DAFF~\cite{Belanger00,Montenegro,nowak}. Near the
thermodynamical phase transition point the universality class of the
transition is determined by several exponents, among them by the
correlation length exponent, ({\it if} the transition is
continuous). The critical percolation phenomenon near the
thermodynamical phase transition point may contribute there and
introduce extra corrections, which have to be taken into account when
the thermodynamical correlation length exponent is
determined.

This paper is organized so that it starts with an introduction of
the random field Ising model in the next section. Also
the numerical method solving exactly the ground states is introduced.
In Section~\ref{perkolaatio_H} the percolation phenomenon is studied,
with a non-zero external field. The universality class of the 
percolation behavior is determined and the dependence of the critical 
external field on the random field strength is investigated.
Section~\ref{perkolaatio_0} concentrates on the percolation phenomenon
without an external field and compares it
with the cases when the external field is applied. The properties
of the spanning cluster are studied in Section~\ref{perc-clust}.
Implications of the percolation to the phase diagram
are discussed, together with the conclusions in 
Section~\ref{concl}.

\section{Random Field Ising Model and Numerical Method}\label{method}

The random field Ising model is defined by its energy Hamiltonian
\begin{equation}
{\mathcal H} = -J \sum_{\langle i j \rangle}S_i S_j - \sum_i 
(h_i + H) S_i,
\label{eq:H}
\end{equation}
where $J>0$ (throughout this paper we set $J=1$, since the relevant value is
its ratio with the random field strength) is the coupling constant
between nearest-neighbor spins $S_i$ and $S_j$. We use here cubic
lattices. $H$ is a constant external field, which if non-zero is
assigned to all of the spins, and $h_i$ is the random field, acting on
each spin $S_i$. We concentrate only on a Gaussian distribution for
the random field values
\begin{equation}
P(h_i) = \frac{1}{\sqrt{2 \pi} \Delta} \exp\left[-\frac{1}{2}
\left(\frac{h_i}{\Delta}\right)^2\right], 
\label{eq:Gaussian}
\end{equation}
with the disorder strength given by $\Delta$ (in this paper $\Delta$
actually denotes the ratio between disorder strength and the coupling
constant), the standard deviation of the distribution. The arguments
presented in this paper could be extended to other lattices and
other distributions, e.g. uniform and bimodal, too. However, discrete
distributions, such as the bimodal one, suffer from degeneracies, and
when calculating thermodynamical quantities
extra averaging, over the degeneracies, has to be done when using discrete
distributions~\cite{Alexander,Sorin}.

To find the ground state structure of the RFIM means that the
Hamiltonian (\ref{eq:H}) is minimized, in which case the positive
ferromagnetic coupling constants prefer to have all the spins aligned
to the same direction. On the other hand the random field contribution
is to have the spins to be parallel with the local field, and thus has
a paramagnetic effect. This competition of ferromagnetic and
paramagnetic effects leads to a complicated energy landscape and the
finding the ground state becomes a global optimization problem.  An
interesting detail of the RFIM is that for $H=0$ it has an experimental
realization as a diluted antiferromagnet in a field. By
gauge-transforming the Hamiltonian of DAFF
\begin{equation}
{\mathcal H} = -J \sum_{\langle i j \rangle}S_i S_j \epsilon_i \epsilon_j 
- B \sum_i \epsilon_i S_i,
\label{eq:DAFF}
\end{equation}
where the coupling constants $J<0$, $\epsilon_i$ is the occupation
probability of a spin $S_i$, and $B$ is now a uniform external field,
one gets the Hamiltonian of RFIM~(\ref{eq:H}) with $H=0$
\cite{Fish79,Cardy,Belanger97}. The ferromagnetic order in
the RFIM corresponds to antiferromagnetic order in the DAFF,
naturally.

For the numerical calculations a graph-theoretical combinatorial
optimization algorithm developed in computer science has been
used. The Hamiltonian (\ref{eq:H}) is transformed to a random flow
graph widely used in computer science with two extra sites: the source
and the sink. The positive field values $h_i$ correspond flow
capacities $c_{it}$ connected to the sink ($t$) from a spin $S_i$,
similarly the negative fields with $c_{is}$ are connected to the
source ($s$), and the coupling constants $2J_{ij} \equiv c_{ij}$
between the spins correspond flow capacities $c_{ij} \equiv c_{ji}$
from a site $S_i$ to its neighboring one $S_j$~\cite{Alavaetal}.  In
the case the external field is applied, only the local sum of fields,
$H+h_i$, is added to a spin toward the direction it is positive. The
algorithms, namely maximum-flow minimum-cut algorithms, enable us to
find the bottleneck, which restricts the amount of the flow which is
possible to get from the source to the sink through the capacities, of
such a random graph.  This bottleneck, path $P$ which divides the
system in two parts: sites connected to the sink and sites connected
to the source, is the global minimum cut of the graph and the sum of
the capacities belonging to the cut $\sum_P c_{ij}$ equals the maximum
flow, and is smaller than of any other path cutting the system. The
value of the maximum flow gives the total minimum energy of the system
and the minimum cut defines the ground state structure of the system,
so that all the spins in the source side of the cut are the spins
pointing down, and the spins in the sink side of the cut point up.
The maximum flow algorithms can be proven to give the exact minimum cut
of all the random graphs, in which the capacities are positive and
with a single source and sink~\cite{network}. We have used a
sophisticated method for solving the maximum flow - minimum cut
problem called push-and-relabel by Goldberg and
Tarjan~\cite{Goltar88}, which we have optimized for our purposes. It
scales almost linearly, ${\mathcal O}(n^{1.2})$, with the number of
spins and gives the ground state in about minute for a million of
spins in a workstation.

We have used periodic boundary conditions in all of the cases. Also
the percolation is tested in the periodical or cylindrical way, i.e.,
a cluster has to meet itself when crossing a boundary in order to span
a system. Finding the spanning cluster has been done using the
usual Hoshen-Kopelman algorithm~\cite{hoshen}.

\section{Percolation with an external field}\label{perkolaatio_H}

\begin{figure}[f]
\centerline{\epsfig{file=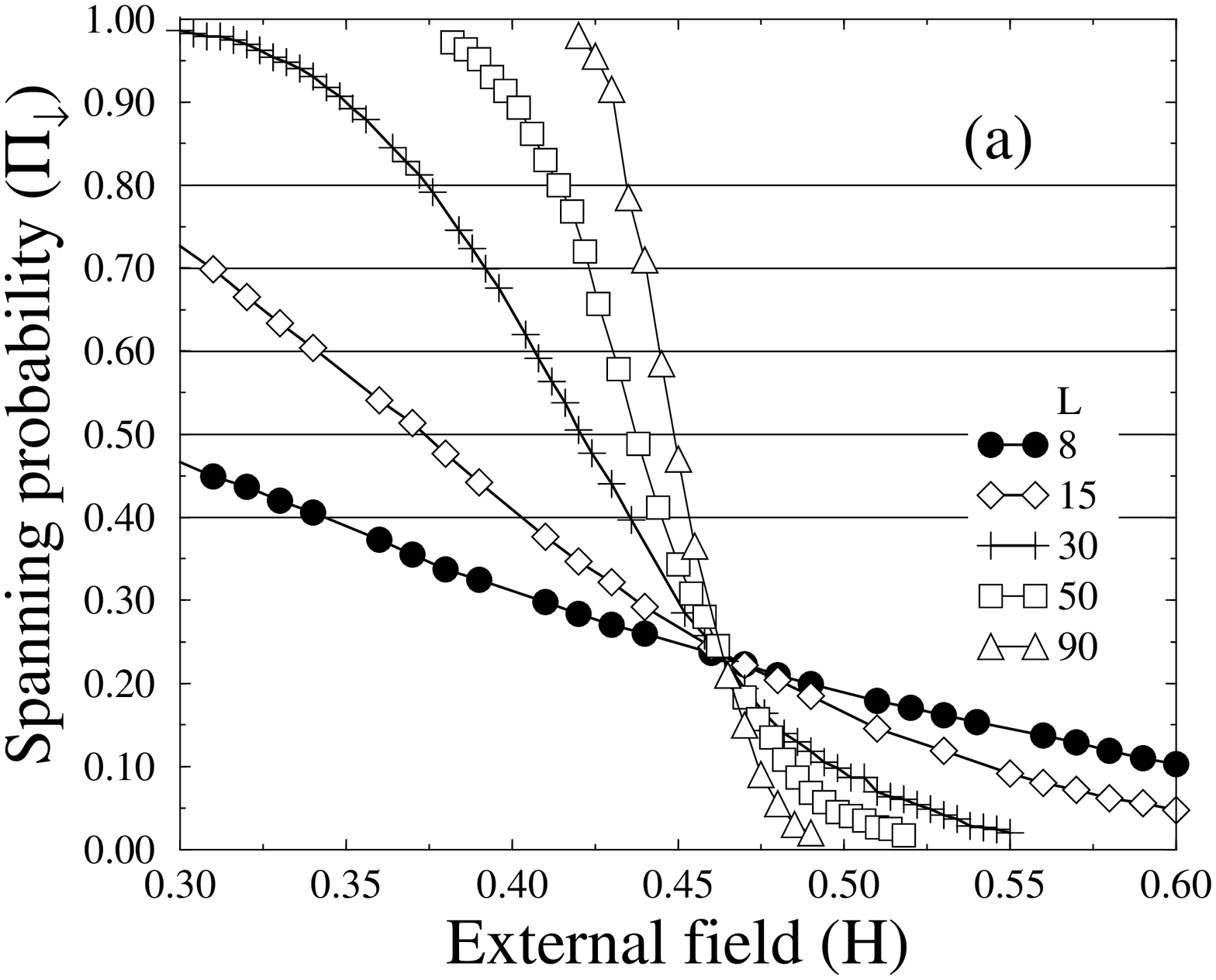,width=7cm,angle=-90}}
\centerline{\epsfig{file=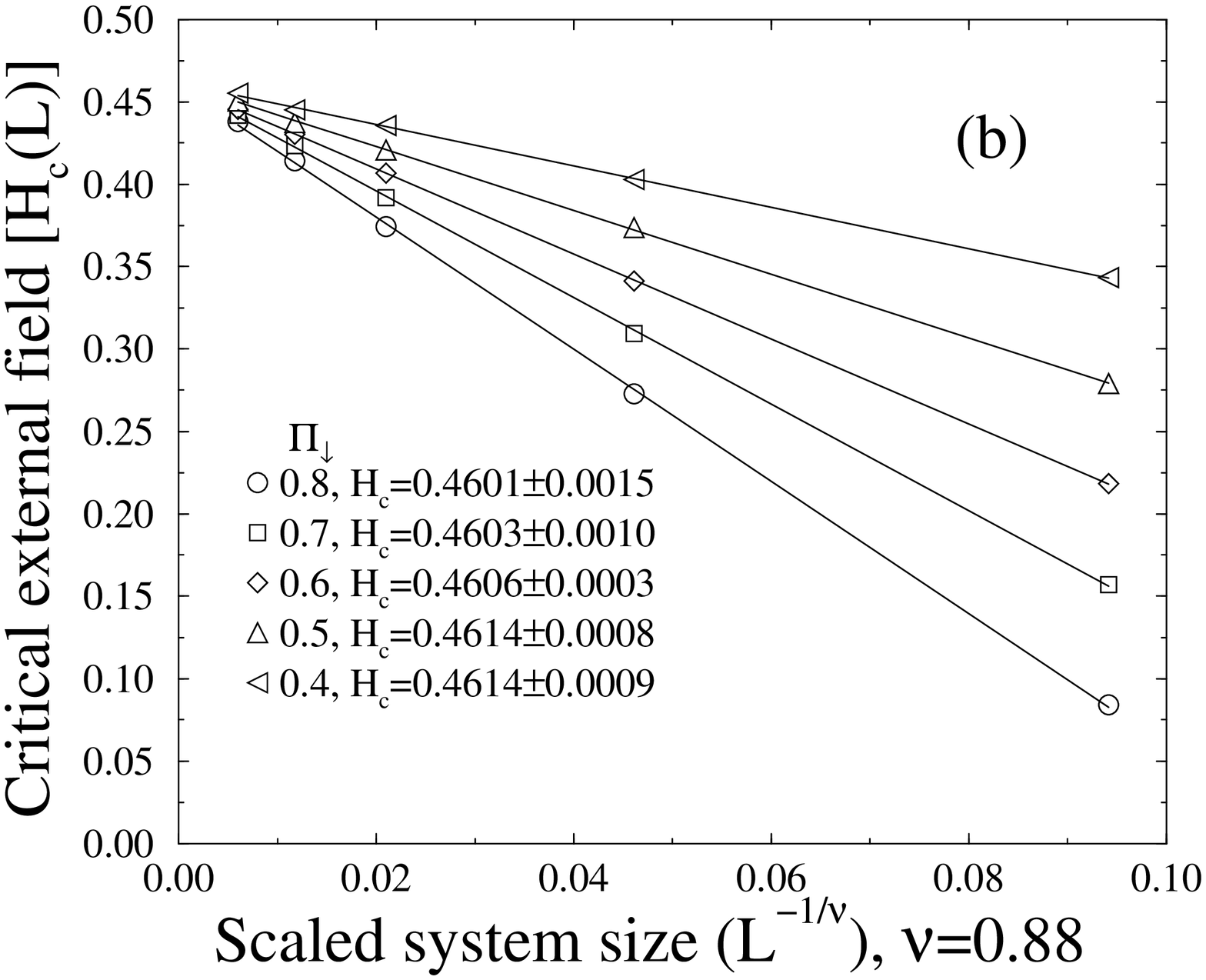,width=7cm,angle=-90}}
\centerline{\epsfig{file=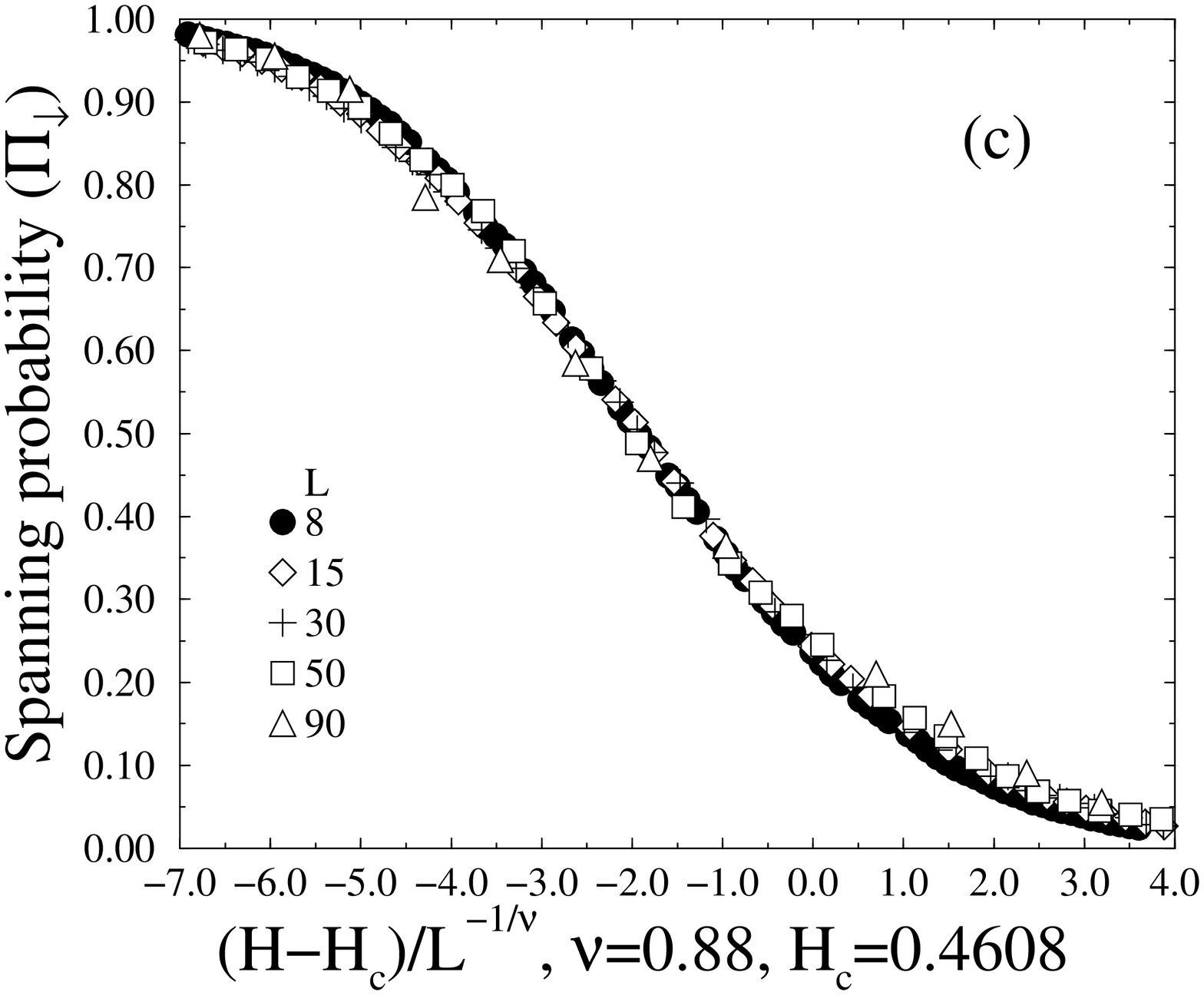,width=7cm,angle=-90}}
\caption{(a) The spanning probabilities of minority down spins 
$\Pi_{\downarrow}$ as a function of upward external field $H$ for
$\Delta =3.5$ with $L^3 \in [8^3\--90^3]$.  The number of realizations
varies between 5000 realizations for $L=8$ and 200 for $L=90$. (b) The
finite size scaling of the fields $H_c(L)$, which are from the
crossing points of the spanning probability curves with the horizontal
lines in (a), leading to the estimate of the critical $H_c = 0.461\pm
0.001$ using $L^{-1/\nu}$, $\nu =0.88$.  The error-bars in the labels
of the figure for different $H_c$ are the errors of the least-squares
fits.  (c) The data-collapse of different system sizes with the
corresponding critical $H_c=0.4608$.}
\label{fig1}
\end{figure}

As a start of the percolation studies of the 3D RFIM we draw in
Fig.~\ref{fig1}(a) the spanning probabilities of down spins
$\Pi_{\downarrow}$ with respect to the uniform external field $H$
pointing up for several system sizes $L$ and for a fixed random field
strength $\Delta = 3.5$. The curves look rather similar to 
standard percolation, except that in site percolation the
systems span at high occupation probability limit, and here the down
spins do not span, when the external positive field has a large value,
and thus the step in the spanning probability is inverse compared to
the one in the occupation percolation. It is interesting to note also,
that since we are using periodic boundary conditions in all of the
directions, also for spanning, the
$\Pi_{\downarrow}(L)$ -lines for various system sizes cross at rather
low $\Pi_{\downarrow}$ values. This is the case for the other $\Delta$,
too. Similar boundary condition dependent behavior have been seen
in the standard percolation, too~\cite{stauffer,lin,ziff}. When we
take the crossing points $H_c(L)$ of the spanning probability curves
with fixed spanning probability values $\Pi_{\downarrow} = $ 0.4, 0.5,
0.6, 0.7, and 0.8, for each systems size $L$, we get an estimate for
the critical external field $H_c$ using finite size scaling, see
Fig.~\ref{fig1}(b). There we have attempted with success to find the
value for $H_c$ using the standard short-range correlated 3D
percolation correlation length exponent $\nu=0.88$~\cite{Amnon}. Using
the estimated $H_c = 0.461 \pm 0.001$ for $\Delta=3.5$ we show a
data-collapse of $\Pi_{\downarrow}$ versus $(H-H_c)/L^{-1/\nu}$ in
Fig.~\ref{fig1}(c), which confirms the estimates of $H_c$ and
$\nu=0.88$. We get similar data-collapses for various other random
field strength values $\Delta$ as well.

Considering the percolation and critical external field with
respect to the random field strength, there is an obvious constraint
in the phase diagram $H$ vs. $\Delta$.  Below the phase transition
critical point, $\Delta_c \simeq 2.27$
\cite{middlefisher,anglesourlas,hartmannowak}, only one of the spin
orientations may span a system, since in a ferromagnetic system the
magnetization has a finite, positive or negative, value and thus there
can not be a massive percolation cluster of the opposite spin
direction. Since the earlier studies of the phase transition at 3D RFIM
\cite{heikopeter,anglesourlas,hartmannowak,sourlas,hartmannyoung,machtaetal}
have shown that the order parameter exponent $\beta$ has a value close
to zero, if not zero, the transition is sharp and therefore the
simultaneous percolation of the both (up {\it and} down) spin
directions should vanish or have vanished at $\Delta_c$ when
approaching from above. The question now remains, whether this takes
place exactly at the phase transition point, so that the critical
points would coincide, or for a $\Delta_p > \Delta_c$. In the latter
case it is also of interest what happens for $H=0$ between the
critical points, on the line $\Delta_c < \Delta < \Delta_p$.  We now
propose a phase diagram, Fig.~\ref{fig2}, for the percolation
phenomenon, and ask at which value the dashed lines in the diagram
meet. Above we showed that in the direction of the vertical arrow at
$H>0$ the universal standard percolation correlation length exponent
is valid. What about at the vertical arrow, what are the critical
exponents there?

To answer the question how the percolation critical external field
$H_c$ behaves with respect to the random field strength $\Delta$, we
have attempted a critical type of scaling using the calculated
$H_c(\Delta)$ for various $\Delta =$2.5, 2.6, 2.75, 3.0, 3.25, 3.5,
4.0, and 4.5.  We have been able to use the Ansatz 
\begin{equation}
H_c \sim (\Delta -\Delta_p)^\delta,
\label{Eq:H_c_D_c}
\end{equation} 
where $\delta=1.31\pm 0.03$ by assuming $\Delta_p = 2.43$, see
Fig.~\ref{fig3}(a).  In Fig.~\ref{fig3}(b) on the other hand we have
plotted the calculated $\Delta$ values versus the scaled critical
external field $[H_c(\Delta)]^{1/1.31}$ and it gives the estimate for
$\Delta_p = 2.43 \pm 0.01$. This indicates that the percolation
probability lines for up and down spins to lose their spanning
property meet at $\Delta_p = 2.43\pm 0.01$.  Note, that our studies in
two-dimensional RFIM gave the values $\Delta_p = 1.65 \pm 0.05$ and
$\delta=2.05\pm 0.10$ for systems {\it to span}, not {\it to lose the
spanning property} as here~\cite{RF2d_perk}.  We also tested various
exponential scaling assumptions for the $H_c(\Delta)$ scaling,
but none of them worked. However, here
we know, that $H_c$ has to vanish at some finite $\Delta_p$ value,
which is greater than or equal to $\Delta_c$.

We have also calculated the order parameter of the
percolation, the probability that a down spin belongs to the
down-spin spanning cluster $P_\infty$. Using the scaling  for
the correlation length
\begin{equation}
\xi_{perc} \sim |H-H_c|^{-\nu},
\label{xi_perc}
\end{equation}
and for the order parameter, when $L < \xi_{perc}$,
\begin{equation}
P_\infty(H) \sim (H_c-H)^{\beta},
\label{beta}
\end{equation}
we get the limiting behaviors,
\begin{equation}
P_\infty(H,L) \sim \left\{ \begin{array}{lll}
(H_c-H)^{\beta} &\mbox{\hspace{5mm}}& L < \xi_{perc},\\
L^{-\beta/\nu}& &L > \xi_{perc}, 
\end{array} \right.
\label{P_infty_limit}
\end{equation}
and thus the scaling behavior for the order parameter becomes
\begin{eqnarray}
P_\infty(H,L) \sim L^{-\beta/\nu} F\left[\frac{(H_c-H)^{-\nu}}{L}\right] 
\nonumber \\
\sim L^{-\beta/\nu} f\left(\frac{H_c-H}{L^{-1/\nu}}\right).
\label{P_infty_scale}
\end{eqnarray}
Note, that here and later in this article $\beta$ denotes the
percolation order parameter exponent as opposed to the bulk phase
transition order parameter exponent discussed earlier in this
paper. We have done successful data-collapses, i.e., plotted the
scaling function $f$, for various $\Delta$ using the standard 3D
short-range correlated percolation exponents $\beta=0.41$ and $\nu
=0.88$, of which the case $\Delta =4.5$ with $H_c=1.0441$ is shown in
Fig.~\ref{fig4}. Note, that only the left part (below zero) of the
scaling function is shown, since $P_\infty(H,L)$ is limited between
[0,1]. When one divides it by $L^{-\beta/\nu}$ the part where
non-scaled $P_\infty(H,L)$ had a value of unity the scaled
$P_\infty(H,L)/L^{-\beta/\nu}$ saturates at different values depending
on $L$.  One can easily see, that the smallest system size $L^3 = 8^3$
does not scale (the rest are scattered around each other and do not
have any trend). We believe that this is due to an intrinsic length
scale over which the spins are correlated, and which depends on the
random field strength value. This will be
discussed in more detail in Section~\ref{perc-clust}, when the scaling
of the spanning cluster is studied.

Hence, we conclude that the percolation transition for a fixed
$\Delta$ versus the external field $H$ is in the standard 3D
short-range correlated percolation universality
class~\cite{Amnon}. This is confirmed by the fractal dimension of the
spanning cluster, too, as discussed below. The fact that the critical
behavior of the percolation with respect to the external field belongs
to the standard short-range correlated percolation universality class
is not surprising, since the strong disorder limit can be seen to be
related with the site percolation problem and that e.g. the
positive external field decreases the number of the occupied down
spins. Also other exponents could be measured, as $\gamma$ for the
average size $\langle s \rangle $ of the clusters, and $\sigma$ and
$\tau$ for the cluster size distribution as well as the fractal
dimension of the backbone of the spanning cluster, the fractal
dimension of the chemical distance, the hull exponent etc.

\section{Percolation at $H=0$}\label{perkolaatio_0}

\begin{figure}[f]
\centerline{\epsfig{file=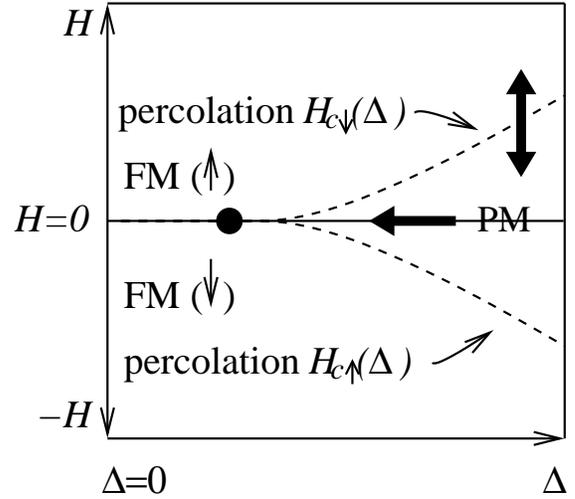,width=8cm}}
\caption{The phase diagram for the minority spin percolation of the 
3D RFIM with disorder strength $\Delta$ and an applied external field
$H$.  The dashed lines define the percolation thresholds $H_c(\Delta)$
for up and down spins to lose their spanning property, below and above
which the minority spins do not percolate anymore. The phase
transition point for the ferro and paramagnetic phases at $\Delta
=2.27$, $H=0$ is shown as a circle.}
\label{fig2}
\end{figure}

\begin{figure}[f]
\centerline{\epsfig{file=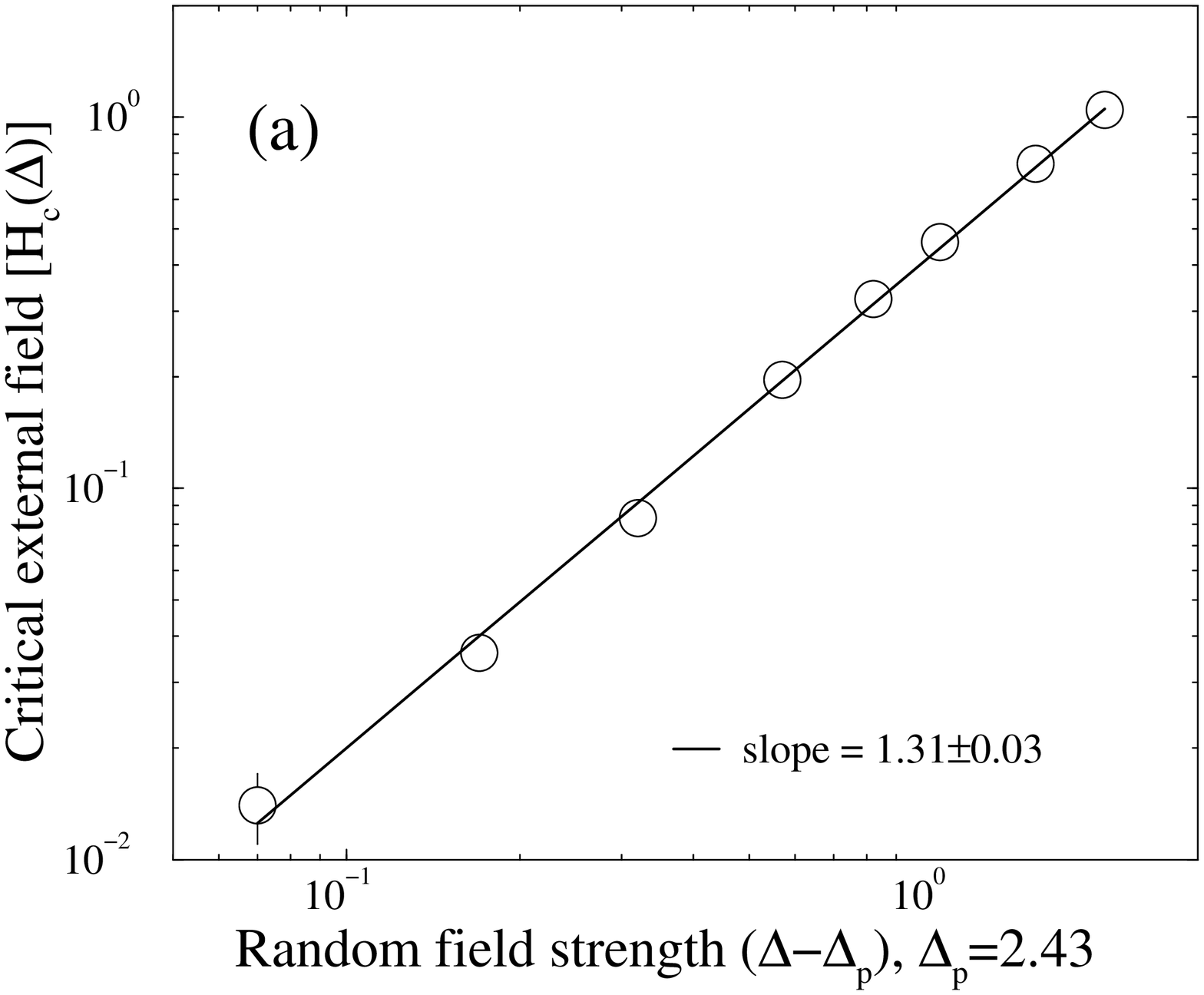,width=7cm,angle=-90}}
\centerline{\epsfig{file=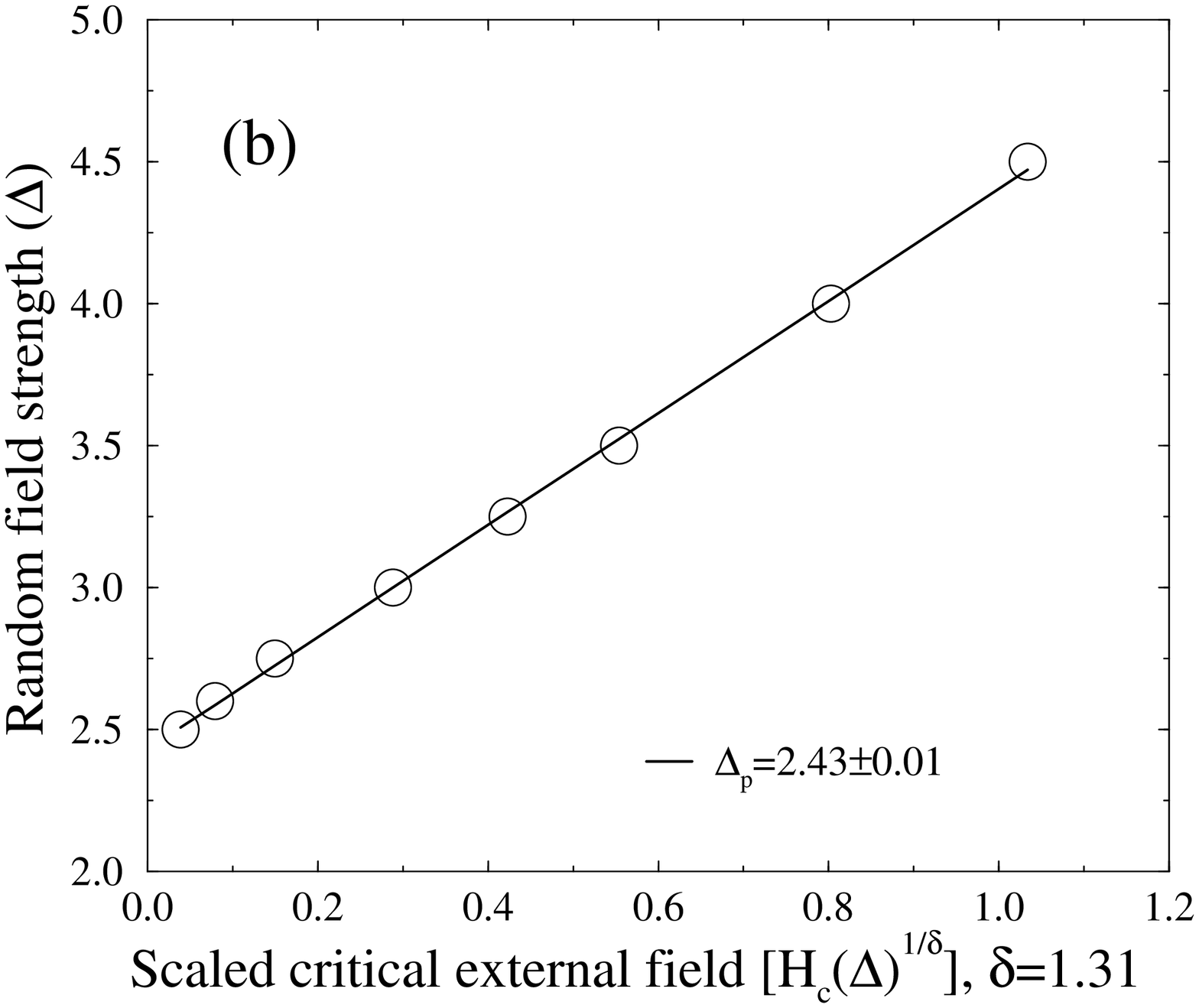,width=7cm,angle=-90}}
\caption{(a) For each calculated $\Delta$ the critical positive
$H_c(\Delta)$ for down spin spanning versus $\Delta -\Delta_p$, where
$\Delta_p$ is estimated to be 2.43. The power law behavior suggests a
scaling: $H_c \sim(\Delta -\Delta_p)^\delta$, where $\delta = 1.31
\pm 0.03$. The error-bar for $\delta$ is the error of the
least-squares fit. (b) The same data but plotted as each $\Delta$
versus $[H_c(\Delta)]^{1/\delta}$, where $\delta =1.31$, which
estimates that at $\Delta_p = 2.43\pm 0.01$ $H_c = 0$. Again the
error-bar for $\Delta_p$ is the error of the least-squares fit. The
other details are as in Fig.~\ref{fig1}.}
\label{fig3}
\end{figure}

\begin{figure}[f]
\centerline{\epsfig{file=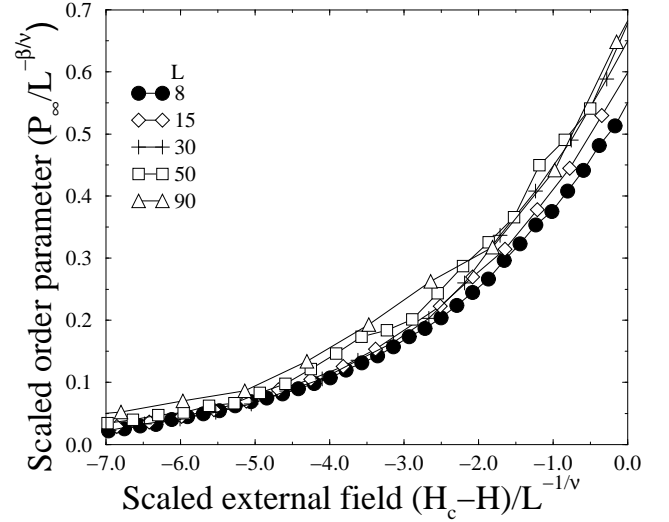,width=7cm,angle=-90}}
\caption{The scaled order parameter, probability that a down spin 
belongs to the down-spin spanning cluster, $P_\infty/L^{-\beta/\nu}$,
$\beta=0.41$, $\nu =0.88$ versus the scaled external field
$(H_c-H)/L^{-1/\nu}$, for $\Delta =4.5$ with $L^3 \in [20^3\--90^3]$.
The data points are disorder averages over 200-5000 realizations.  The
corresponding critical $H_c(\Delta=4.5)=1.0441$.}
\label{fig4}
\end{figure}

In the previous section we learned that the dashed lines at the phase
diagram, Fig.~\ref{fig2}, meet at the value $\Delta_p = 2.43 \pm
0.01$, which is well above the phase transition critical point
$\Delta_c = 2.27$. This raises the question, how this is seen, when
the external field $H=0$ and what happens between $\Delta_c$ and
$\Delta_p$. Thus we study the phase diagram in the direction of the
horizontal arrow in Fig.~\ref{fig2}. There are two strategies
for this that we employ separately to evaluate their advantages
and disadvantages. That is, one can take the $\Delta_p$ to
be a priori the same for all $\Pi_{\uparrow \downarrow}$,
the probability for simultaneous spanning of up {\it and} down spins.
Or then this can be let to vary with $\Pi_{\uparrow \downarrow}$,
as in two-dimensions~\cite{RF2d_perk}.

\begin{figure}[f]
\centerline{\epsfig{file=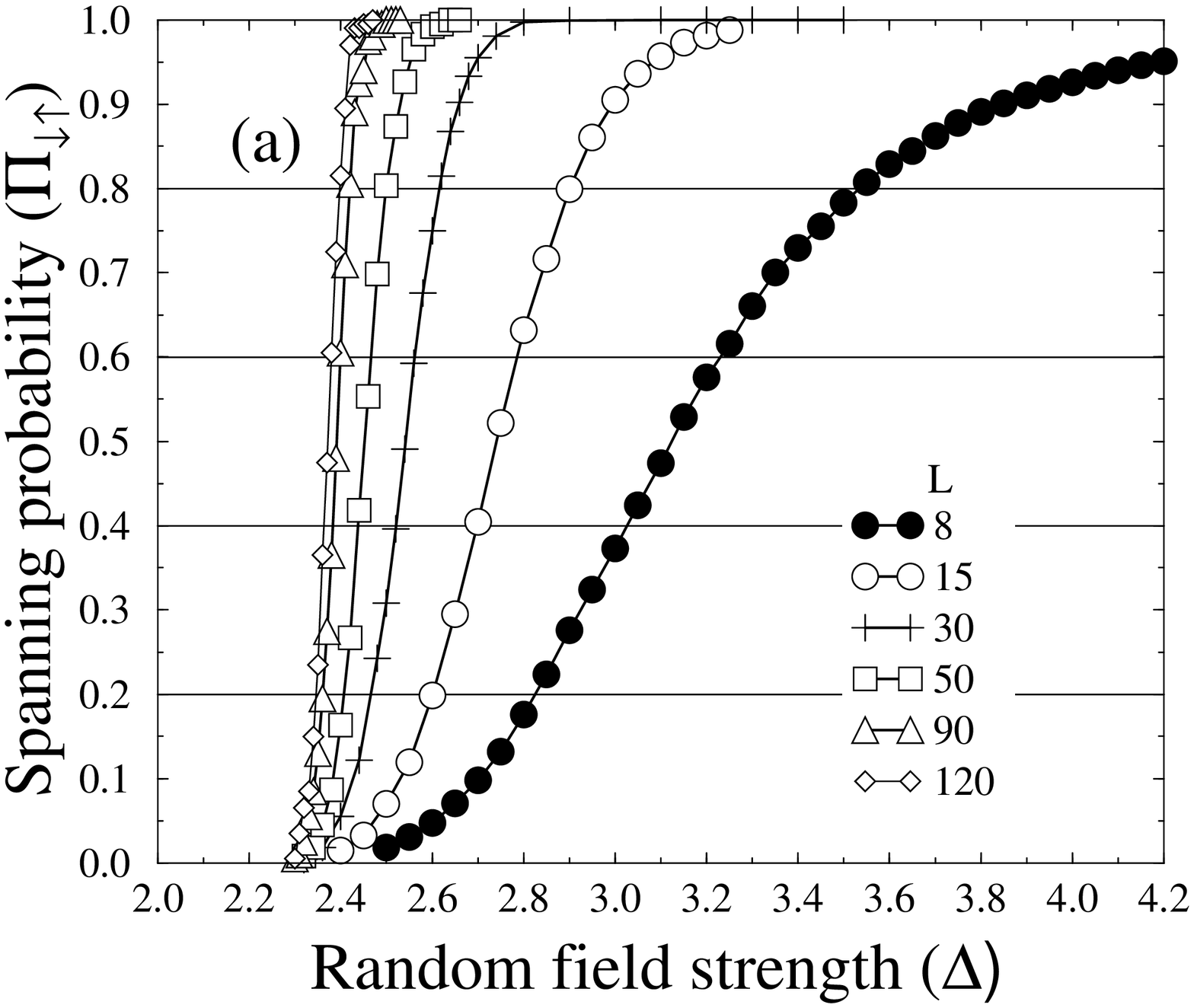,width=7cm,angle=-90}}
\centerline{\epsfig{file=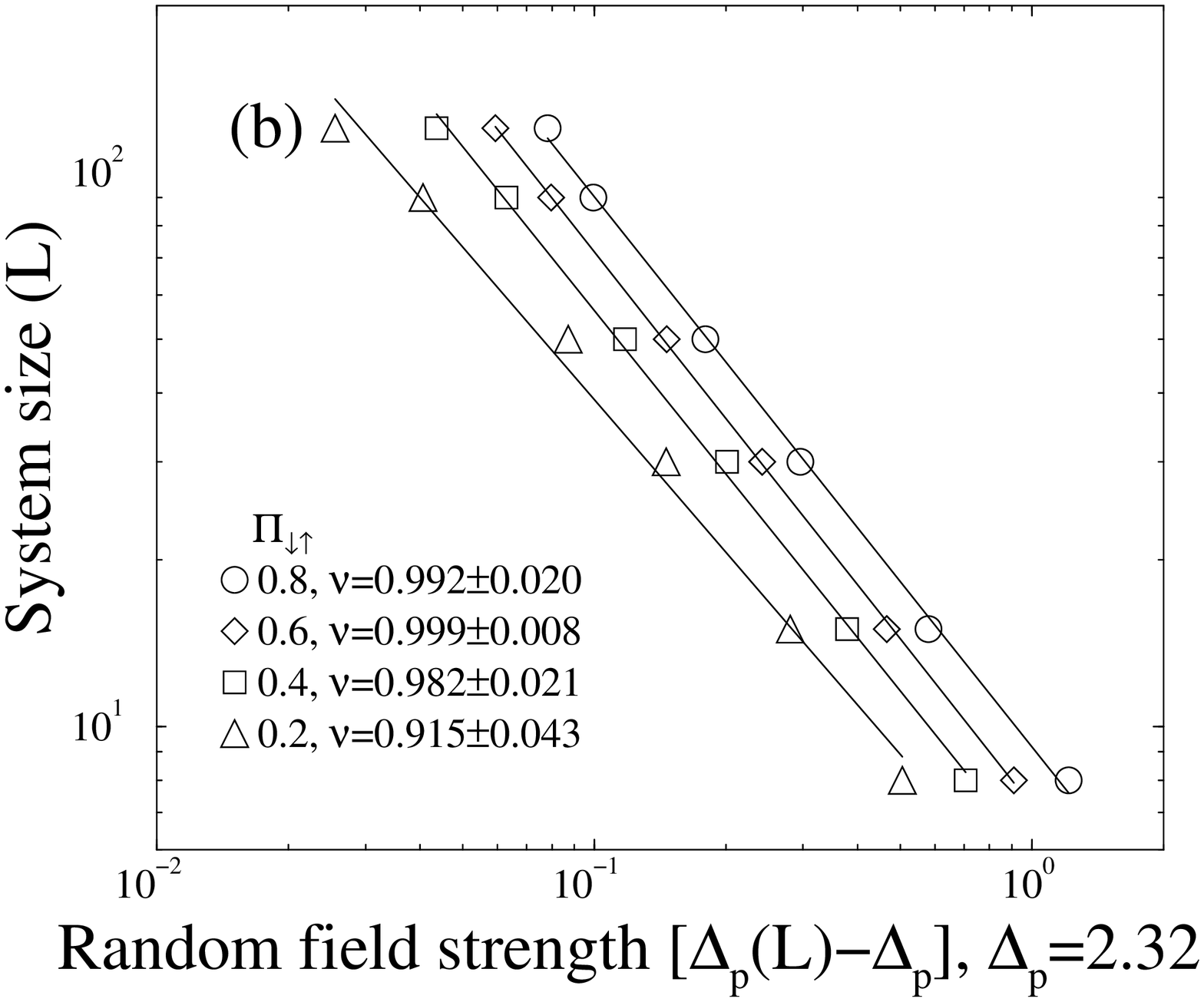,width=7cm,angle=-90}}
\centerline{\epsfig{file=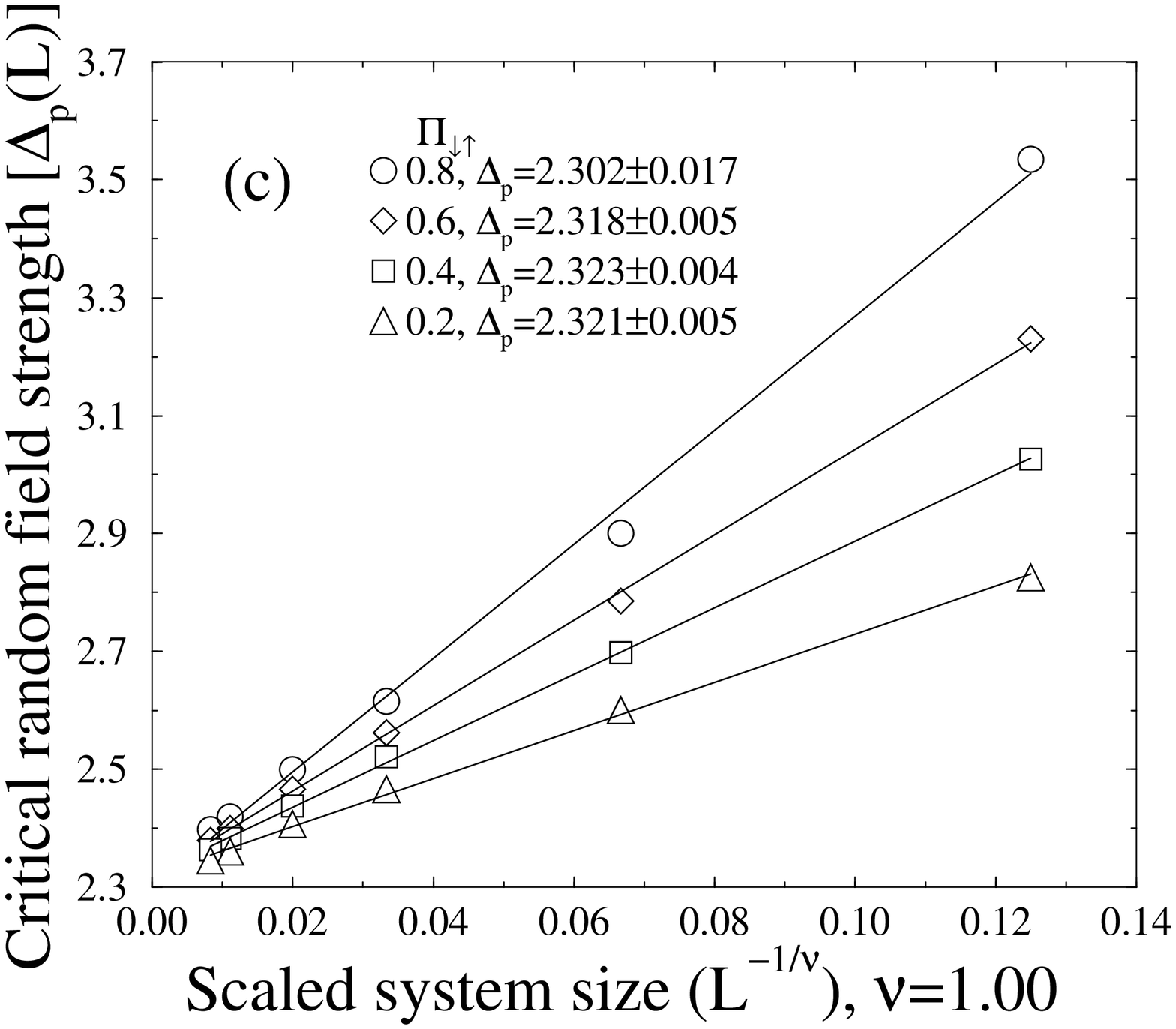,width=7cm,angle=-90}}
\centerline{\epsfig{file=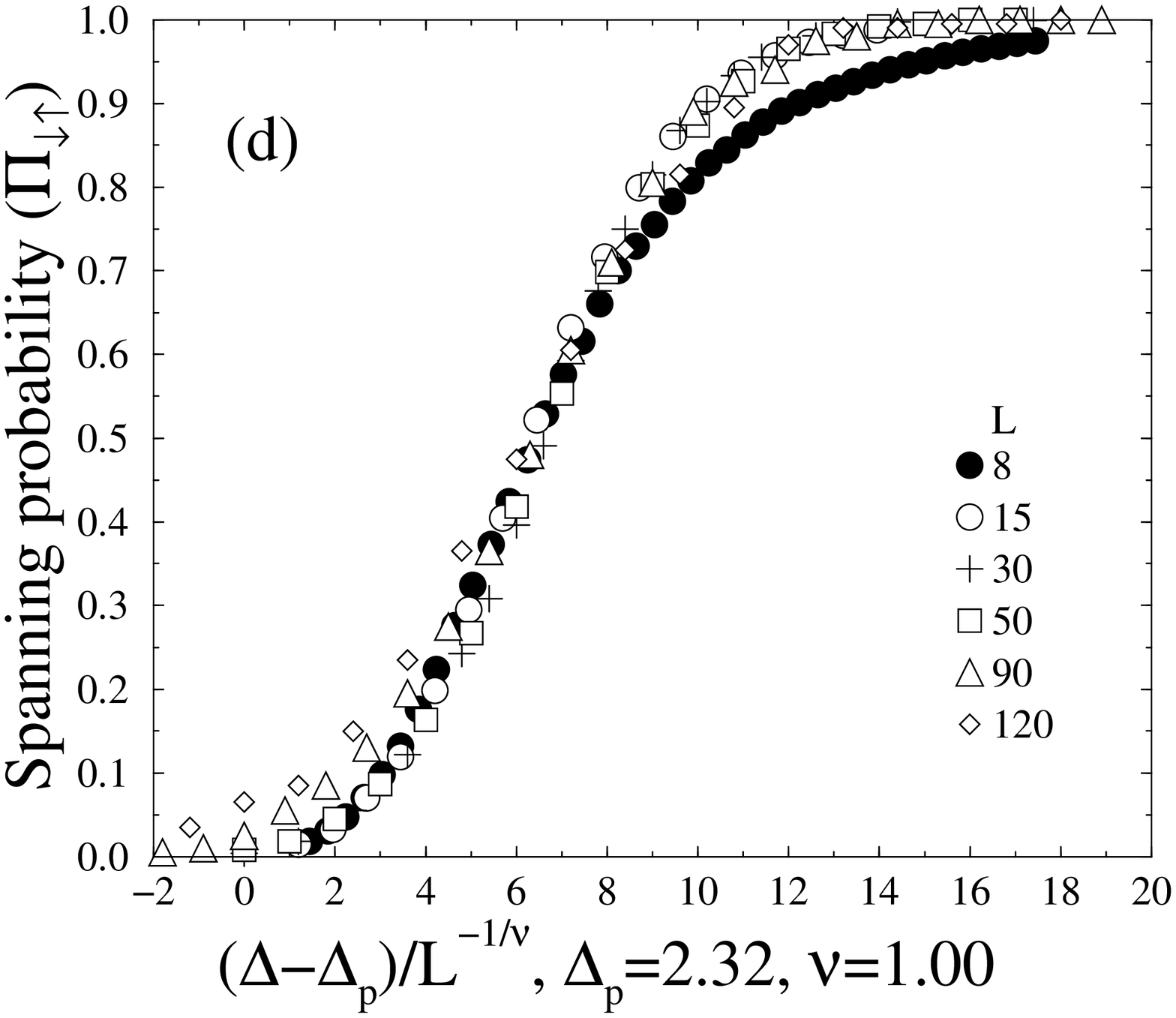,width=7cm,angle=-90}}
\caption{(a) The spanning probabilities for system sizes 
$L^3 \in [8^3\--120^3]$ of simultaneous up and down spin spanning
$\Pi_{\uparrow \downarrow}$ as a function of $\Delta$ for $H=0$. The
data points are disorder averages over 200-5000 realizations. (b) Each
system size $L$ versus $\Delta_p(L) -\Delta_p$, where $\Delta_p(L)$'s
are the corresponding crossing points of the spanning probability
curves with the horizontal lines of $\Pi_{\uparrow \downarrow}=$ 0.2,
0.4, 0.6, and 0.8 in (a) and $\Delta_p$ is estimated to be 2.32. The
power law behavior suggests a scaling: $L \sim(\Delta
-\Delta_p)^{-\nu}$, where $\nu = 0.97 \pm 0.05$. The error-bars in the
labels of the figure for different $\nu$'s are the errors of the
least-squares fits.  (c) The same data as in (b), but now plotted as
random field strength values $\Delta_p(L)$ versus the scaled system
size $L^{-1/\nu}$, where $\nu =1.0$, leading to a same estimate of
$\Delta_p = 2.32\pm 0.01$. The error-bars in the labels of the figure
for different $\Delta_p$ are the errors of the least-squares fits. (c)
The data-collapse of (a) with the corresponding critical $\Delta_p =
2.32$ and $\nu =1.0$.}
\label{fig5}
\end{figure}

In Fig.~\ref{fig5}(a) we have plotted the probability for simultaneous
spanning of up {\it and} down spins $\Pi_{\uparrow \downarrow}$ as a
function of the Gaussian random field strength $\Delta$ for various
system sizes $L^3=8^3$, $15^3$, $30^3$, $50^3$, $90^3$, and
$120^3$. This case now resembles the standard occupation percolation
in the sense, that the step in the percolation probability is from a
low value to a large value when $\Delta$ is increased.  By
estimating that the $\Delta_{p, H=0}$ at the thermodynamic limit has a
value of 2.32 using fixed $\Pi_{\uparrow \downarrow}=$ 0.2, 0.4, 0.6,
and 0.8 for the $\Delta_{p, H=0}(L)$ we find that the effective $\nu$
gets a value of $0.97 \pm 0.05$ when approaching the critical point in
this direction, see Fig.~\ref{fig5}(b). On the other hand assuming
that the $\nu=1.0$ the $\Delta_{p, H=0}$ becomes $2.32 \pm 0.01$, see
Fig.~\ref{fig5}(c).  These plots show that the estimates should be
correct. However, the data-collapse, Fig.~\ref{fig5}(d), using the
estimates above could be better. Obviously the smallest system size,
$L^3=8^3$, does not scale.

There are a couple of points one should note from the scaling. Firstly
the estimate for the $\Delta_{p, H=0} = 2.32 \pm 0.01$ is still above
the phase transition point $\Delta_c =2.27$.  Another point is that
$\Delta_{p, H=0}$ is reasonably far away from $\Delta_p = 2.43 \pm
0.01$ [note, that the error-bar in the finite field case is the
error-bar of the least-squares fit in Fig.~\ref{fig3}(b) and does
not take into account other sources for the error, e.g., the error of
$\delta$, statistics etc., and thus is a lower limit]. The third point
is that $\Pi_{\uparrow \downarrow} = 0.0$ at $\Delta_{p, H=0}$ and
$\Pi_{\downarrow} = 0.25$ at $H_c(\Delta)$ [for $\Delta=3.5$ see
Fig.~\ref{fig1}(c)].  Our take on the two different estimates is that
they are compatible with the following scenario. For $\Delta$ values
that are slightly below 2.43 one can have {\em only one critical}
spanning cluster, and the probability for this is then
$\Pi_{\downarrow}$, about 0.25. The both orientations do span
simultaneously, as they can do for all $\Delta$-values above
$\Delta_c$, but they should not be both critical, unless one
decreases the disorder strength further.

For the estimate of the correlation length exponent, deviations from
normal percolation are seen since $\nu=1.00\pm 0.05$ instead of
$\nu=0.88$. In our opinion this reflects the fact that for $H \neq 0$
the correlations from the proximity of $\Delta_c$ are negligible,
whereas here the spin-spin correlations change with system size. The
correlation length exponent is higher than that for percolation, so
clear-cut percolation scaling can not be expected.  Differences
between the $H=0$ and $H\neq0$ -cases were found also in the two
dimensional case~\cite{RF2d_perk}. Note, that in two dimension, the
exponent was {\it dependent on the spanning probability} and the
standard correlation length exponent was found where the spanning
probability for either of the spin directions to span $\Pi_{\uparrow /
\downarrow}$ had a non-zero value (remember, that in two dimensional
square lattices without an external field at large $\Delta$ neither of
the spin directions span, and with small $\Delta$ either of them start
to span). 

Here we tried, as in two-dimensions, to do fits using
several criteria for $\Pi_{\uparrow \downarrow}=$ [0.05, 0.15, 0.20,
\ldots, 0.95] and letting both $\Delta_p$ and $\nu$ vary depending on
$\Pi_{\uparrow \downarrow}$. Indeed, we obtained monotonous behaviors
depending on $\Pi_{\uparrow \downarrow}$ for both $\nu$ and
$\Delta_p$.  However, this may just reflect how finite size effects
depend on the criterion. It is anyhow worth of noting that for
$\Pi_{\uparrow \downarrow}$ approaching zero, $\Delta_p$ gets 
also closer and closer to 2.27, i.e., 
the accepted value for the phase transition point $\Delta_c$.
Moreover the correlation length exponent moves towards
$\nu =1.3\pm 0.1$, in the neighborhood of
the phase transition correlation length
exponents reported in the
literature~\cite{middlefisher,hartmannowak,hartmannyoung}.  Similarly
if $\Pi_{\uparrow \downarrow}$ is let to approach unity,
$\Delta_p$ closes on the value $\Delta_p=2.43$ obtained above,
in the finite field case. This behavior may be just coincidence, or
related to the ($\Delta$-dependent) correlations in the system, to
how they change the universality class of percolation in the vicinity
of the phase transition. We return to this 
in the conclusions, in Section~\ref{concl}.

Hence we have shown that at large $\Delta$ both of the spin directions
span simultaneously, and by decreasing random field strength we find a
critical $\Delta_{p, H=0}$, which is above the phase transition point
$\Delta_c$, and below which there is no simultaneous
spanning. Therefore we conclude, that in the whole regime $\Delta_c <
\Delta \leq \Delta_{p, H=0}$ there is geometrical criticality in 3D RF magnets,
since always only either of the spin directions spans the
system. However, the spanning cluster cannot be massive there, i.e.,
scale with the Euclidean dimension ($d=3$), the system still being
paramagnetic, but has to be a fractal. The scaling of the spanning
cluster is studied in the next section and the implications of the
critical region in Section~\ref{concl}.

\section{The spanning cluster}\label{perc-clust}

\begin{figure}[f]
\centerline{\epsfig{file=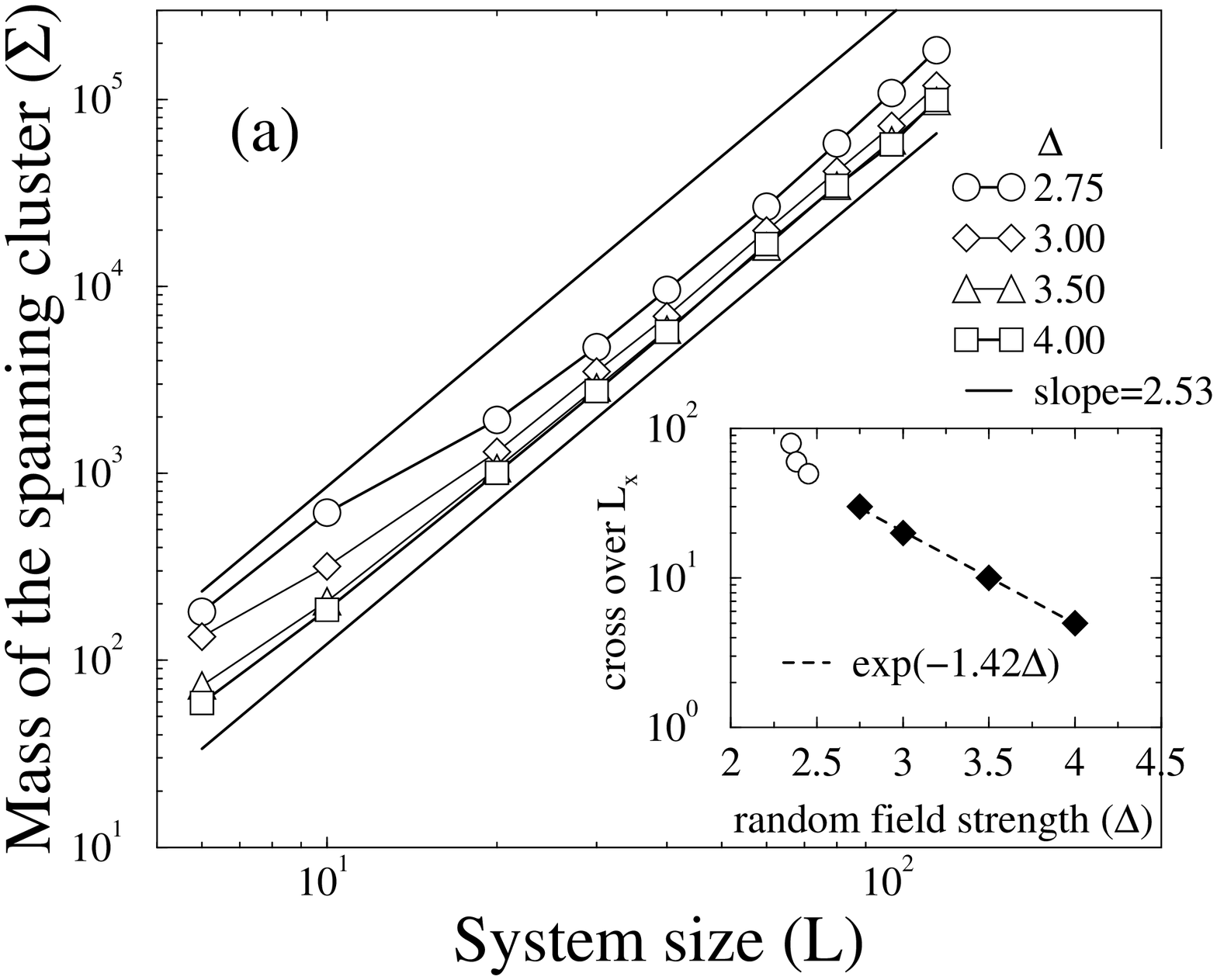,width=7cm,angle=-90}}
\centerline{\epsfig{file=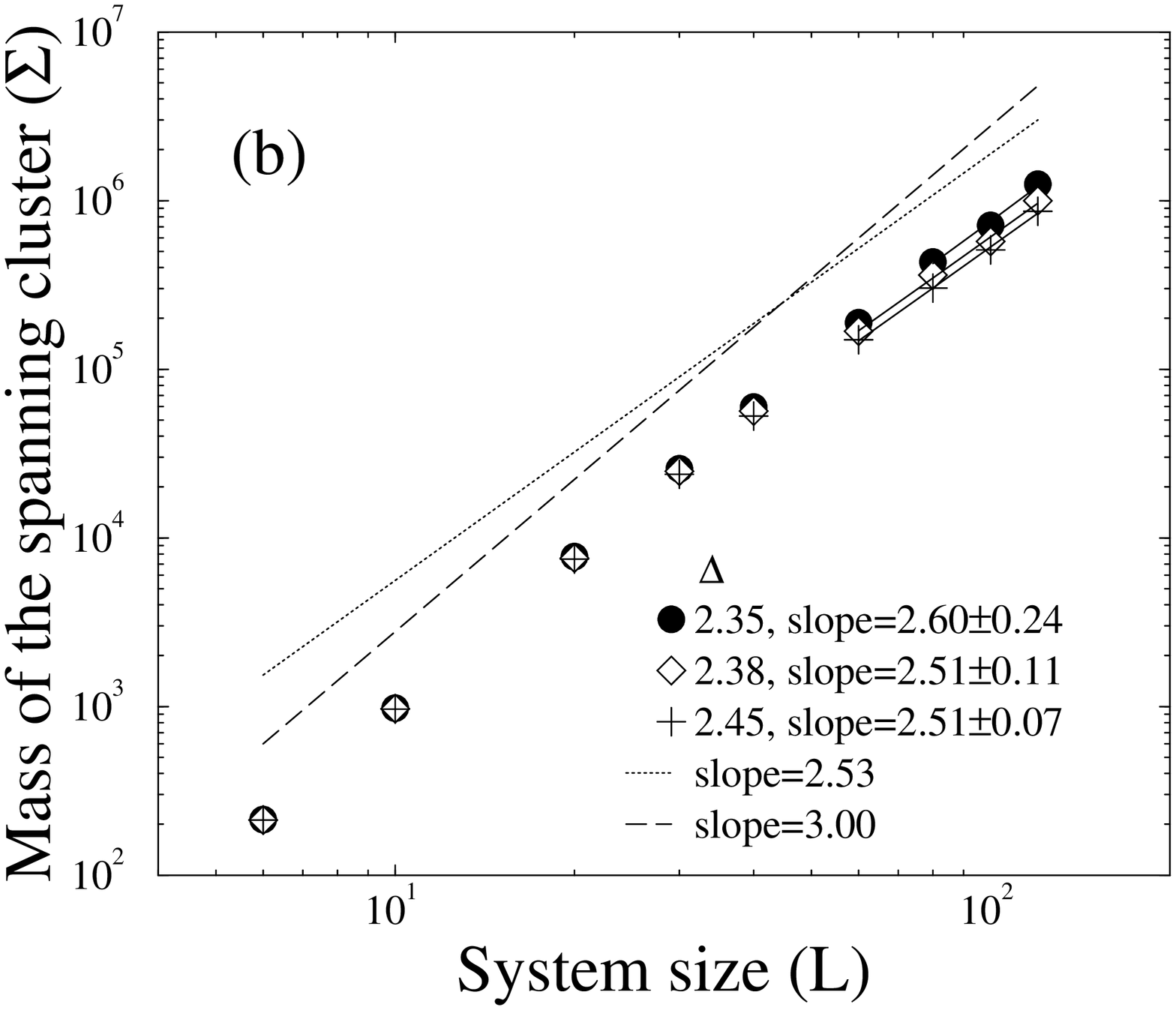,width=7cm,angle=-90}}
\caption{(a) The average mass of spanning cluster of down spins 
for random field strength values $\Delta=$ 2.75, 3.0, 3.5, and 4.0 at
the critical positive external field value $H_c(\Delta)$ [see the
values from Fig.~\ref{fig3}(b)]. The 3D percolation fractal dimension
$D_f = 2.53$ is indicated with solid lines.  In the inset a crossover
length scale at which system size the asymptotic behavior is met for
each random field strength is plotted as closed diamonds. The
least-squares fit estimates an exponential behavior with a slope of
$-1.42 \pm 0.03$. (b) The average mass of spanning clusters of either
spin orientations for random field strength values $\Delta=$ 2.35,
2.38, and 2.45 when $H=0$. The solid lines are the least-squares fits
to the data with the slopes indicated in the labels. The dotted line
with a slope of $D_f =2.53$ and the dashed line with a slope of $d=3$
are guides to eye.  The estimated crossover length scales are plotted
in the inset of Fig.~\ref{fig6}(a) as open circles.}
\label{fig6}
\end{figure}

In Fig.~\ref{fig6}(a) we have plotted the mass of the spanning cluster
of down spins with respect to the system size at
$H_c(\Delta)>0$ for four random field strength values $\Delta= $ 2.75,
3.0, 3.5, and 4.0 up to the system size $L^3 =120^3$. As a guide to
the eye the fractal dimension $D_f = 2.53$ of the standard percolation
is drawn in the figure and the systems can be seen asymptotically
approaching the same scaling. However, there are
obvious finite size effects, which depend on $\Delta$. We have
estimated roughly the crossover system sizes for the systems to reach
the correct scaling, $L_x \simeq 30$, 20, 10, and 5 for $\Delta= $
2.75, 3.0, 3.5, and 4.0, respectively. This hints about an exponential
scaling with a slope of $-1.42 \pm 0.03$ for the crossover length
scale, see closed diamonds in the inset of Fig.~\ref{fig6}(a).  
The above scaling predicts for $\Delta
=4.5$ $L_x \simeq 3$, smaller than $L=8$ (in Fig.~\ref{fig4}, this size
does not scale) but note that the
prefactors of the scaling behaviors need not to be the same.
In Fig.~\ref{fig6}(b) we have drawn for three $\Delta \leq
\Delta_p$, i.e., $\Delta=$2.35, 2.38, and 2.45 (which is so close to
$\Delta_p$ that its $H_c$ is practically zero with respect to the 
numerical precision, $10^{-3}$) at $H=0$ the scaling of the mass of
the spanning cluster of either of the spin orientations up to system
size $L^3 =120^3$.  There one can see that the fractal dimension $D_f
= 2.53$ of the standard percolation is asymptotically met, too, but at
much larger system sizes.  Here we have estimated the crossover system
sizes $L_x \simeq 80$, 60, and 50, for $\Delta=$2.35, 2.38, and 2.45,
respectively. They are plotted as open circles in the inset of
Fig.~\ref{fig6}(a) and are obviously diverging from the exponential
behavior mentioned above when approaching phase transition
$\Delta_c$. These large values for $L_x$ do not leave much room for
the asymptotic scaling, since it is difficult to go above
$L^3 =120^3$. However, the crossover is visible. There is one another
thing one notices from Figs.~\ref{fig6}(a) and (b). In the case we plot
the mass of the spanning cluster of the down spins in $\Delta >
\Delta_p$ and $H_c(\Delta) >0$ the crossover is from a smaller slope
to the asymptotic $D_f = 2.53$ one.  In the case $\Delta \leq
\Delta_p$ the crossover is from the Euclidian dimension (slope of
three, i.e., effective ferromagnetism) to the asymptotic $D_f = 2.53$.
There it is obviously affected by the vicinity of the phase transition
point.

\section{Discussion and conclusions}\label{concl}

In this paper we have studied the character of the ground state of the
three-dimensional random field Ising magnet in, mostly, the
paramagnetic phase.  A geometrical critical phenomenon exists in these
systems: for cubic lattices in ordinary percolation both occupied
and unoccupied sites span the systems, when the occupation probability
is one half. In the RFIM this corresponds to the case with a high random 
field strength value, without an external field. When an external field is
applied and the random field strength decreased, a percolation
transition, for the other spin orientation to lose the spanning
property, can be seen. The transition is shown to be in the standard 3D
short-range correlated percolation universality class, when studied as
a function of the external field. Hence, the correlations in the
three-dimensional random field Ising magnets are only of finite extent
as could be expected in this region of the bulk phase diagram. Based on
our numerical results both the critical points $\pm H_c (\Delta)$ approach
when $\Delta$ is decreased, and finally meet at a $\Delta_p \simeq 2.43 > 
\Delta_c$, at which $H_c=0$.
When the percolation transition is studied without an external field and
tuning the random field strength similar behavior is
found, i.e. signatures of a percolation line (a $\Delta_p > \Delta_c$).
This might cause puzzling consequences when studying the
character of the ground states, because the percolation correlations
may influence the magnetization correlation length.

The major theoretical implications have to do with the phase
transition. Note that earlier groundstate studies of the domain structure
implied that there is only a ``one-domain state'' below the critical
field, and a ``two-domain state'' in the paramagnetic phase (extending
down from high disorder values) \cite{Esser}.
If the transition is first-order, then one expects the
percolation properties of the paramagnetic phase to be discontinuous in
the thermodynamic limit. If the transition is second-order, then one
may ask what is the correct way to link the presence of the
percolation transition to the critical phase?  At $\Delta_c$, one
expects that the spin-spin correlations show power-law
correlations. For a normal percolation transition, these are (as in
the disordered phase in general) of short-range character. 
There is a divergent length scale as the
transition is approached from the paramagnetic phase, below which the
spin-spin correlations matter and the scaling of the spanning
clusters is volume-like. 

Assume that the properties of the largest cluster are governed by
the power-law correlations. An old result by Weinrib gives a Harris'
criterion for this approach, to check how this would change its
structure from ordinary percolation~\cite{weinrib}.  If the site
occupation probability correlations decay as $r^{-a}$, one has that
the decay is relevant if $a \nu_{old} - 2 < 0 \rightarrow \nu_{new} =
2/a$, where now $\nu_{old} = 0.88$ for 3D site percolation. One gets a
critical decay exponent $a_c = 2.27$, much larger than that found by
Middelton and Fisher \cite{middlefisher}, which is very close to
zero. An application of the theory of correlated percolation would
thus imply that the spin-spin correlations at $\Delta_c$ are relevant
for percolation.
They would change the universality class, of percolation, in a way that
would reflect such correlations. This conclusion should be taken with
plenty of salt, obviously.

One should note also that although this study was done using cubic
lattices it can be extended to other lattices, too, since all the
common three dimensional lattices have $p_c <0.5$. Thus the
transition from the both spin orientations spanning phase to only one
spin orientation spanning phase should exists. In the case of diluted
antiferromagnets the percolation is already seen as percolation of
diluted spins. The implication of this paper is that the influence of
percolation is even more rich. Lately there has been interest in
studying domain walls and excitations in RF magnets. In both cases the
underlying percolation criticality should affect the structure of the
clusters that result from varying the boundary conditions.

\section*{Acknowledgments}

This work has been supported by the Academy of Finland Centre of
Excellence Programme. It was also performed under the auspices of the
U.S.\ Dept.\ of Energy at the University of California/Lawrence
Livermore National Laboratory under contract no. W-7405-Eng-48 (ETS).



\end{multicols}
\widetext

\end{document}